\documentclass[twocolumn,letter]{jpsj3}
\usepackage{graphicx}
\usepackage{color}
\usepackage{bm}

\title{%
Superconducting Diode Effect due to the Inverse Spin Hall Effect
in Josephson Junctions with Extrinsic Spin-Orbit Interaction
}

\author{%
Yositake Takane$^{1,*}$, Aur\'{e}lien Manchon$^{2}$, and Gen Tatara$^{3}$
}

\inst{%
$^{1}$Graduate School of Advanced Science and Engineering,
Hiroshima University, Higashihiroshima, Hiroshima 739-8530, Japan\\
$^{2}$Aix-Marseille Universit\'{e}, CNRS, CINaM, 13288 Marseille Cedex 09,
France\\
$^{3}$RIKEN Center for Emergent Matter Science (CEMS), 2-1 Hirosawa, Wako, Saitama 351-0198, Japan
}

\recdate{ \hspace{50mm} }

\abst{%
The spin density and Josephson current at equilibrium are studied numerically
in a superconductor-normal metal-superconductor (SNS) junction
with extrinsic spin-orbit interaction in the N segment.
This system preserves structural inversion symmetry.
We find that the Josephson current generates the spin Hall effect, 
resulting in an opposite spin density near the two edges of the N segment.
We also find that a Zeeman magnetic field with gradient applied
to the N segment generates an anomalous phase shift due to
the inverse spin Hall effect, resulting in a diode effect
when higher harmonics of the Josephson current are present.
These results are consistent with those of a previous theoretical study
based on diagrammatic perturbation theory.
We find that the diode effect becomes more pronounced
when stronger coupling between the N and S segments
enhances the higher harmonics.
}

\begin{document}

\maketitle

The critical current of a superconducting system can vary depending on
the flow direction under certain conditions.
The direction-dependent critical current is referred to as
the superconducting diode effect.
While this effect was initially considered for
Josephson junctions,~\cite{Hu,Reynoso1,Reynoso2,Yokoyama1,Yokoyama2}
it was first observed in a superconductor with a superlattice structure
that breaks inversion symmetry.~\cite{Ando}
Since then, the diode effect has been demonstrated experimentally in
various superconductors~\cite{Lin,Bauriedl,Hou,Le,Wan} and Josephson
junctions,~\cite{Pal,Wu,Baumgartner,Costa,Reinhardt,Lotfizadeh,Zhang,Jeon}
in which time-reversal and spatial inversion symmetries are broken.
Theoretical studies have also been conducted on corresponding
systems.~\cite{Daido,He,Ilic,Yuan,Cayao,Kokkeler,Yerin,Mori,Sahoo}

Theoretical and experimental studies on the diode effect have mainly
focused on superconducting systems that break structural inversion symmetry,
such as those with Rashba spin-orbit interaction (SOI).
However, this does not mean that the structural inversion symmetry breaking
is essential for this effect.
A recent theoretical study~\cite{Tatara} based on
diagrammatic perturbation theory described a scenario for the diode effect
in a superconductor-normal metal-superconductor (SNS) junction
with extrinsic SOI due to random impurities in the N segment.
This system preserves the structural inversion symmetry.~\cite{comment1}.
After demonstrating that the Josephson current induces
an opposite spin density near the two edges of the N segment
(i.e., the spin Hall effect),~\cite{Bergeret}
the theoretical study shows that a spatially inhomogeneous magnetic field
generates an anomalous phase shift due to the inverse spin Hall effect,
which results in the diode effect in the presence of higher harmonics,
$I_{n}\sin(n\chi)$ with $n \ge 2$, of the Josephson current,~\cite{Reinhardt}
where $\chi$ represents the phase difference between the two S segments.
Since an inhomogeneous magnetic field  extrinsically breaks
the inversion symmetry, the diode effect appears
without the structural inversion symmetry breaking.
An anomalous phase shift due to a uniform magnetic field
has been pointed out in literature.~\cite{Reynoso1,Reynoso2,Yokoyama1,
Yokoyama2,Bezuglyi,Krive,Buzdin,Zazunov,Liu,Konschelle,Assouline}

\begin{figure}[btp]
\begin{center}
\includegraphics[height=2.0cm]{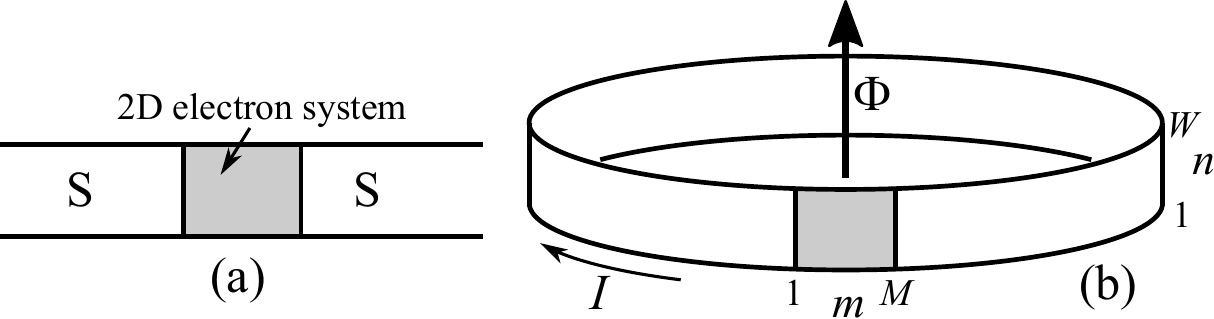}
\end{center}
\caption{
Schematics of (a) a Josephson junction consisting of a 2DES (shaded)
located between two $s$-wave superconductors
and (b) a ring-shaped system with a magnetic flux $\Phi$ piercing the ring.
}
\end{figure}
In this letter, we numerically study the superconducting diode effect
in an SNS junction [Fig.~1(a)]
in which the N segment consists of a two-dimensional electron system (2DES)
with extrinsic SOI due to random impurities.
Focusing on the scenario in Ref.~\citen{Tatara},
we examine the spin density and Josephson current in two regimes:
the tunneling regime, where the coupling between the N and S segments is weak,
and the metallic regime, where the coupling is strong. 
The scenario is confirmed
in the tunneling regime, on which the theoretical study is based.
We find that the diode effect is highly pronounced in the metallic regime,
reflecting an enhancement of higher harmonics of the Josephson current.

As a model of the Josephson junction, we consider a ring-shaped system
[Fig.~1(b)] of width $W$ consisting of the 2DES of length $M$
and the superconductor of length $N$.
In this system with a magnetic flux of $\Phi$,
the circulating persistent current depends on the phase $\varphi$,
which is defined as $\varphi = 2\pi (\Phi/\Phi_{0})$,
where $\Phi_{0}=\frac{h}{2e}$.
As long as the length of the superconductor is much larger
than its coherence length,
the properties of the persistent current are equivalent to those of
the Josephson current in the ordinary SNS model [Fig.~1(a)],
where $\varphi$ corresponds to $\chi$ in the Josephson junction.
Since the ring-shaped system is closed, various physical quantities can be
more easily calculated than in the ordinary model,
in which the two S segments are infinitely long.
Hereafter, the persistent current is referred to as the Josephson
current.

We express the Hamiltonian for the ring-shaped system as
$H = H_{\rm N} + H_{\rm SOI} + H_{\rm S} + H_{\rm T}$,
where $H_{\rm N}$, $H_{\rm SOI}$, $H_{\rm S}$, and $H_{\rm T}$ describe
the unperturbed 2DES, the SOI in the 2DES, the superconductor,
and the coupling between the two subsystems, respectively.
We use $m$ and $n$ to specify the $x$- and $y$-coordinates of the sites
in the system.
Let $c_{m,n\sigma}^{\dagger}$ and $c_{m,n\sigma}$ be the creation
and annihilation operators, respectively, of the electron with spin $\sigma$
at the $(m,n)$th site.
It is convenient to define $C_{m,n}$ as
\begin{align}
   C_{m,n}
   = \bigl[c_{m,n\uparrow}\, c_{m,n\downarrow}\,
           c_{m,n\uparrow}^{\dagger}\, c_{m,n\downarrow}^{\dagger}
     \bigr]^\mathsf{T} .
\end{align}
The Hamiltonian for the 2DES
in the region $1 \le m \le M$ and $1 \le n \le W$ is
\begin{align}
     \label{eq:def-H_N}
  H_{\rm N}
 & = \frac{1}{2}\sum_{m,n}
     \Bigl[ C_{m,n}^{\dagger}h_{0}^{\rm N}C_{m,n}
     + \left( C_{m+1,n}^{\dagger}h_{1}^{\rm N}C_{m,n} + {\rm h.c.} \right)
       \nonumber \\
 & \hspace{12mm}
   + \left( C_{m,n+1}^{\dagger}h_{1}^{\rm N}C_{m,n} + {\rm h.c.} \right)
     \Bigr] ,
\end{align}
where
\begin{align}
  h_{0}^{\rm N} = \left[ \begin{array}{cc}
                           -\mu_{\rm N}\hat{1} & \hat{0} \\
                           \hat{0} & \mu_{\rm N}\hat{1}
                         \end{array}
                  \right] ,
     \hspace{2mm}
  h_{1}^{\rm N} = \left[ \begin{array}{cc}
                           -t \hat{1} & \hat{0} \\
                           \hat{0} & t \hat{1}
                         \end{array}
                  \right] .
\end{align}
Here, $\hat{1}$ and $\hat{0}$ are $2 \times 2$ unit and zero matrices,
respectively.
The Hamiltonian for the superconductor in the region
$M+1 \le m \le M+N$ and $1 \le n \le W$ is
\begin{align}
     \label{eq:def-H_S}
  H_{\rm S}
 & = \frac{1}{2}\sum_{m,n}
     \Bigl[ C_{m,n}^{\dagger}h_{0}^{\rm S}C_{m,n}
   + \left( C_{m+1,n}^{\dagger}h_{1}^{\rm S}C_{m,n} + {\rm h.c.} \right)
       \nonumber \\
 & \hspace{12mm}
   + \left( C_{m,n+1}^{\dagger}h_{1}^{\rm S}C_{m,n} + {\rm h.c.} \right)
     \Bigr] ,
\end{align}
where
\begin{align}
  h_{0}^{\rm S} = \left[ \begin{array}{cc}
                           -\mu_{\rm S}\hat{1} & i\Delta\sigma_{y} \\
                           -i\Delta\sigma_{y} & \mu_{\rm S}\hat{1}
                         \end{array}
                  \right] ,
     \hspace{2mm}
  h_{1}^{\rm S} = \left[ \begin{array}{cc}
                           -t \hat{1} & \hat{0} \\
                           \hat{0} & t \hat{1}
                         \end{array}
                  \right] .
\end{align}
Here and hereafter, $\sigma_{x}$, $\sigma_{y}$, and $\sigma_{z}$ represent
Pauli matrices.
The transfer integral in the superconductor is assumed to be the same as
that in the 2DES.
Instead of determining the pair potential $\Delta$ in a self-consistent manner,
we assume it to be a constant in the superconductor.
This is a good approximation for analyzing the Josephson current.
The coupling between the 2DES and the superconductor is described by
\begin{align}
     \label{eq:def-H_T}
  H_{\rm T}
   = \frac{1}{2}\sum_{n}
     \Bigl[ \left( C_{1,n}^{\dagger}h_{1}^{\rm T}C_{M+N,n}
                   +C_{M+1,n}^{\dagger}h_{2}^{\rm T}C_{M,n} \right)
                   + {\rm h.c.}
     \Bigr] ,
\end{align}
where
\begin{align}
  h_{1}^{\rm T} = \left[ \begin{array}{cc}
                           -t'\hat{1} & \hat{0} \\
                           \hat{0} & t'\hat{1}
                         \end{array}
                  \right] ,
     \hspace{2mm}
  h_{2}^{\rm T} = \left[ \begin{array}{cc}
                           -t' e^{-i\frac{\varphi}{2}}\hat{1} & \hat{0} \\
                           \hat{0} & t' e^{i\frac{\varphi}{2}}\hat{1}
                         \end{array}
                  \right] .
\end{align}
The effects of $\Phi$ are included through $e^{\pm i\frac{\varphi}{2}}$.
We assume that the SOI arises from attractive impurities in the 2DES,
with each impurity located at the center of a square plaquette.
Let $\bm{R}_{i}=(m_{i}+\frac{1}{2},n_{i}+\frac{1}{2})a$ be the position of
the $i$th impurity, where $a$ is the lattice constant.
We adopt the following model for the SOI:
\begin{align}
     \label{eq:def-H_SOI}
  H_{\rm SOI}
 & = \frac{1}{2}\sum_{i}\lambda_{i}
     \Bigl[ i\bigl( C_{m_{i},n_{i}+1}^{\dagger}h_{z}C_{m_{i},n_{i}}
       \nonumber \\
 & \hspace{0mm}
                 + C_{m_{i}+1,n_{i}+1}^{\dagger}h_{z}C_{m_{i},n_{i}+1}
                 - C_{m_{i}+1,n_{i}+1}^{\dagger}h_{z}C_{m_{i}+1,n_{i}}
       \nonumber \\
 & \hspace{0mm}
                 - C_{m_{i}+1,n_{i}}^{\dagger}h_{z}C_{m_{i},n_{i}}
            \bigr)
          + {\rm h.c.}
     \Bigr] ,
\end{align}
where $\lambda_{i}>0$ represents the strength of SOI
due to the $i$th attractive impurity, and
\begin{align}
  h_{z} = \left[ \begin{array}{cc}
                    \sigma_{z} & \hat{0} \\
                    \hat{0} & \sigma_{z}
                         \end{array}
                  \right] .
\end{align}
We treat $\{\lambda_{i}\}$ as a set of random variables.
The Fourier representation of $H_{\rm SOI}$ is expressed as
\begin{align}
     \label{eq:def-H_SOI_k}
  H_{\rm SOI}
 & = \frac{1}{2}\sum_{i}\lambda_{i}\frac{1}{A}\sum_{\bm{k},\bm{k}'}
     e^{-i(\bm{k}'-\bm{k})\cdot \bm{R}_{i}}
        \nonumber \\
 & \hspace{-7mm}
   \times
   8\Biggl( \cos\frac{k'_{y}a}{2}\cos\frac{k_{x}a}{2}
            \sin\frac{k'_{x}a}{2}\sin\frac{k_{y}a}{2}
        \nonumber \\
 & \hspace{-1mm}
          - \cos\frac{k'_{x}a}{2}\cos\frac{k_{y}a}{2}
            \sin\frac{k'_{y}a}{2}\sin\frac{k_{x}a}{2}
    \Biggr)
    C_{\bm{k'}}^{\dagger}h_{z}C_{\bm{k}} ,
\end{align}
where $A$ is the area of the 2DES, and $C_{\bm{k}}$ and $C_{\bm{k}}^{\dagger}$
are Fourier transforms of $C_{m,n}$ and $C_{m,n}^{\dagger}$, respectively.
In the long wavelength limit, $H_{\rm SOI}$ reproduces the $\sigma_{z}$ term
of the conventional SOI model, which is proportional to
$ie^{-i(\bm{k}'-\bm{k})\cdot\bm{R}_{i}}(\bm{k}'\times\bm{k})_{z}\sigma_{z}$.

We focus on the spin density in the 2DES and the Josephson current.
The spin density is characterized by
\begin{align}
     \label{eq:S_z^B/T}
  S_{z}^{\rm B} = \sum_{m=1}^{M} \sum_{n=1}^{\frac{W}{2}}
                  \langle S_{m,n}^{z} \rangle ,
     \hspace{5mm}
  S_{z}^{\rm T} = \sum_{m=1}^{M} \sum_{n=\frac{W}{2}+1}^{W}
                  \langle S_{m,n}^{z} \rangle ,
\end{align}
which represent the $z$-component of the spin density
in the bottom and top half regions of the 2DES, respectively.
Here, $\langle S_{m,n}^{z} \rangle$ represents the expectation value of
\begin{align}
  S_{m,n}^{z} = \frac{\hbar}{4}
               C_{m,n}^{\dagger}
               \left( \begin{array}{cc}
                        \sigma_{z} & \hat{0} \\
                        \hat{0} & -\sigma_{z}
                      \end{array} \right) C_{m,n}
\end{align}
in the ground state at zero temperature
and is obtained by rewriting the operators in it
in terms of the bogolon operators, which diagonalize the Hamiltonian $H$.
The Josephson current $I$ is expressed inside the 2DES as
\begin{align}
 I = \frac{i e}{\hbar}\sum_{n = 1}^{W}
     \langle C_{m+1,n}^{\dagger} I_{m,n}C_{m,n}- {\rm h.c.}\rangle
\end{align}
for $1 \le m \le M-1$ with
\begin{align}
  I_{m,n} = \frac{1}{2}
          \left[ \begin{array}{cc}
                   t\hat{1} +i\lambda'\sigma_{z} & \hat{0} \\
                   \hat{0} & t\hat{1} -i\lambda'\sigma_{z}
                 \end{array}
          \right] ,
\end{align}
where $\lambda'$ represents the strength of the SOI
on the link between the $(m,n)$th and $(m+1,n)$th sites.
Here, $I > 0$ when the Josephson current flows
in the direction indicated by the arrow in Fig.~1(b).

To generate the inverse spin Hall effect, we apply a spatially antisymmetric
Zeeman magnetic field with gradient that polarizes the spin density in the 2DES
such that $\langle S_{m,n}^{z} \rangle < 0$ near the bottom edge ($n =1$)
and $\langle S_{m,n}^{z} \rangle > 0$ near the top edge ($n =W$).
Such a magnetic field is described by the spin-dependent potential,
\begin{align}
  V_{m,n} = \left[ \begin{array}{cc}
                     V_{m,n}^{z}\sigma_{z} & \hat{0} \\
                     \hat{0} &  -V_{m,n}^{z}\sigma_{z}
                   \end{array}
            \right] ,
\end{align}
which is added to $h_{0}^{\rm N}$, where
\begin{align}
  V_{m,n}^{z}
  = \left\{ 
      \begin{array}{cc}
       \frac{w+1-n}{w}B & 1 \le n \le w  \vspace{1mm} \\
       -\frac{n-W+w}{w}B & W+1-w \le n \le W
      \end{array}
    \right.
\end{align}
in the region $\frac{M}{8}+1 \le m \le \frac{7M}{8}$,
and $V_{m,n}^{z} = 0$ otherwise.
This means that $|V_{m,n}^{z}| = B$ at the edge sites
in the specified region
and decreases linearly to zero into the interior over width $w$.

We calculate $S_{z}^{\rm T}$, $S_{z}^{\rm B}$, and $I$
as a function of $\varphi$ in the system of $M = 40$, $N = 160$, $W = 40$,
and $w = 5$ with $\Delta = 0.2t$, and $\mu_{\rm N} = \mu_{\rm S} = -3.5t$.
The Fermi wavelength and the coherence length are
given as $\lambda_{\rm F} \approx 8.9a$ and $\xi \approx 7.1a$, respectively.
We put an attractive impurity on each plaquette in the 2DES
with probability $0.2$ and randomly set $\{\lambda_{i}\}$ such that they are
uniformly distributed in the range $(0, \lambda)$ with $\lambda = 0.2t$.
We separately consider the tunneling regime with $t' = 0.3t$
and the metallic regime with $t' = t$.
In the tunneling regime,
the third and higher harmonics of the Josephson current are negligible
and is consistent with the assumption of the scenario.~\cite{Tatara}
The numerical results show $S_{z}^{\rm B} \approx -S_{z}^{\rm T}$
in the absence or presence of the antisymmetric Zeeman field.
The equality $S_{z}^{\rm B} = - S_{z}^{\rm T}$ holds exactly
if the spin density in the superconductor is included in them.

Let us first consider the tunneling regime with $t'=0.3t$
with a single configuration of impurities.
Figure~2 shows (a) $S_{z}^{\rm T}(\varphi)$ in the cases $B/t = 0.0$ and $0.2$
and (b) $I(\varphi)$ in the cases $B/t = 0.0$, $0.1$, and $0.2$.
Here and hereafter, $S_{z}^{\rm T}$ is normalized by $\frac{\hbar}{2}$
and $I(\varphi)$ is normalized by
$I_{0} \equiv (\frac{t'}{t})^4\frac{e \Delta}{\hbar}$.
In the case $B/t =0.0$, $S_{z}^{\rm T}(\varphi)$ satisfies
$S_{z}^{\rm T}(\varphi) = -S_{z}^{\rm T}(-\varphi)$
in accordance with the antisymmetric feature of $I(\varphi)$.
In addition, the spin density satisfies
$S_{z}^{\rm B}(\varphi) \approx -S_{z}^{\rm T}(\varphi)$ as noted above.
These observations show that the Josephson current generates
the spin Hall effect.
The numerical results of $I$ in the case $B/t = 0.0$
are accurately fitted by the trial function
\begin{align}
  I(\varphi) = I_{1}\sin(\varphi) + \delta I_{2}\sin(2\varphi)
\end{align}
with $I_{1} \approx 0.4742$ and
$\delta I_{2} \approx -0.005269$.
\begin{figure}[btp]
\vspace{0mm}
\begin{tabular}{cc}
\begin{minipage}{0.5\hsize}
\begin{center}
\includegraphics[height=4.5cm]{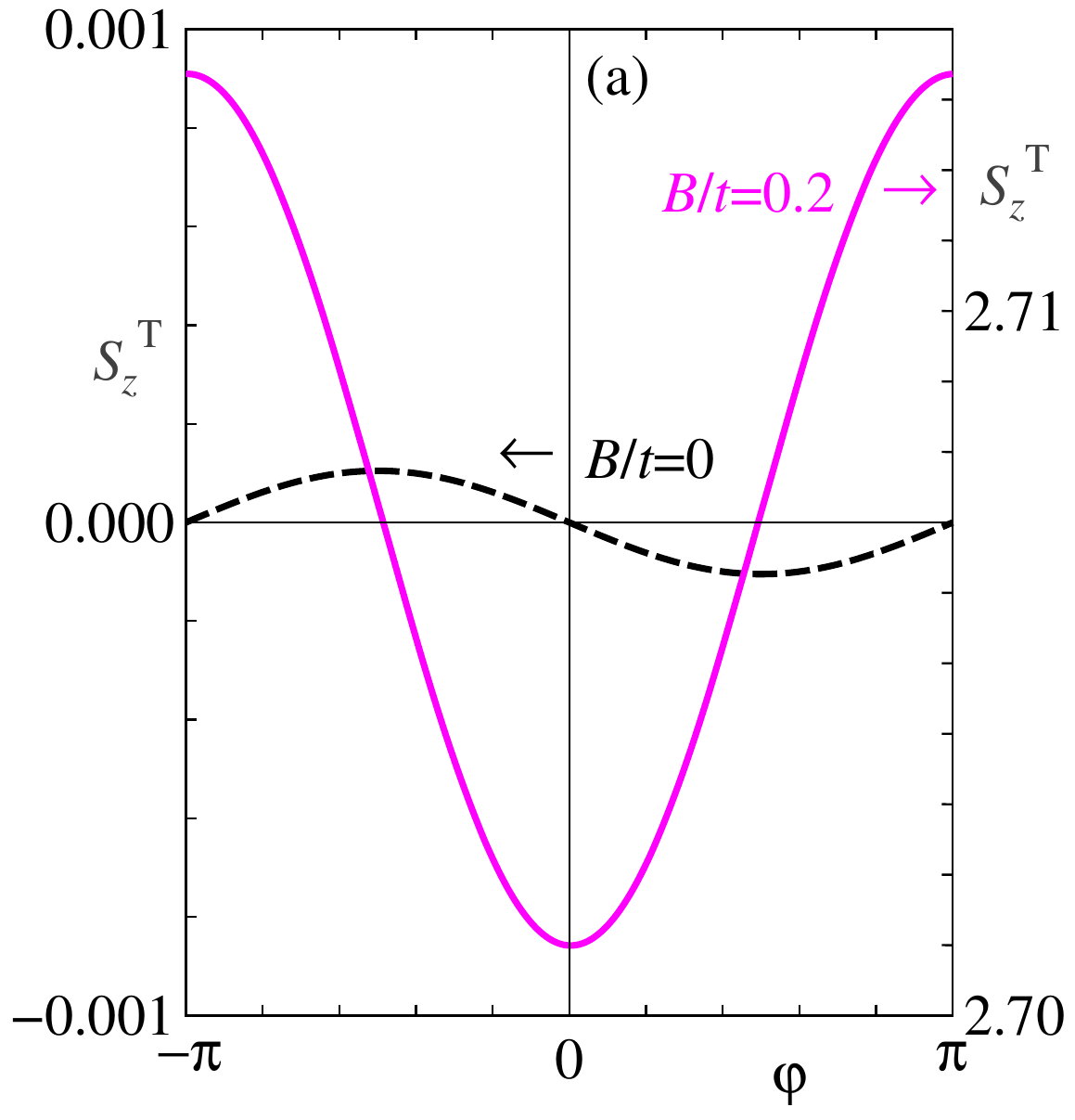}
\end{center}
\end{minipage}
\begin{minipage}{0.5\hsize}
\begin{center}
\hspace{0mm}
\includegraphics[height=4.5cm]{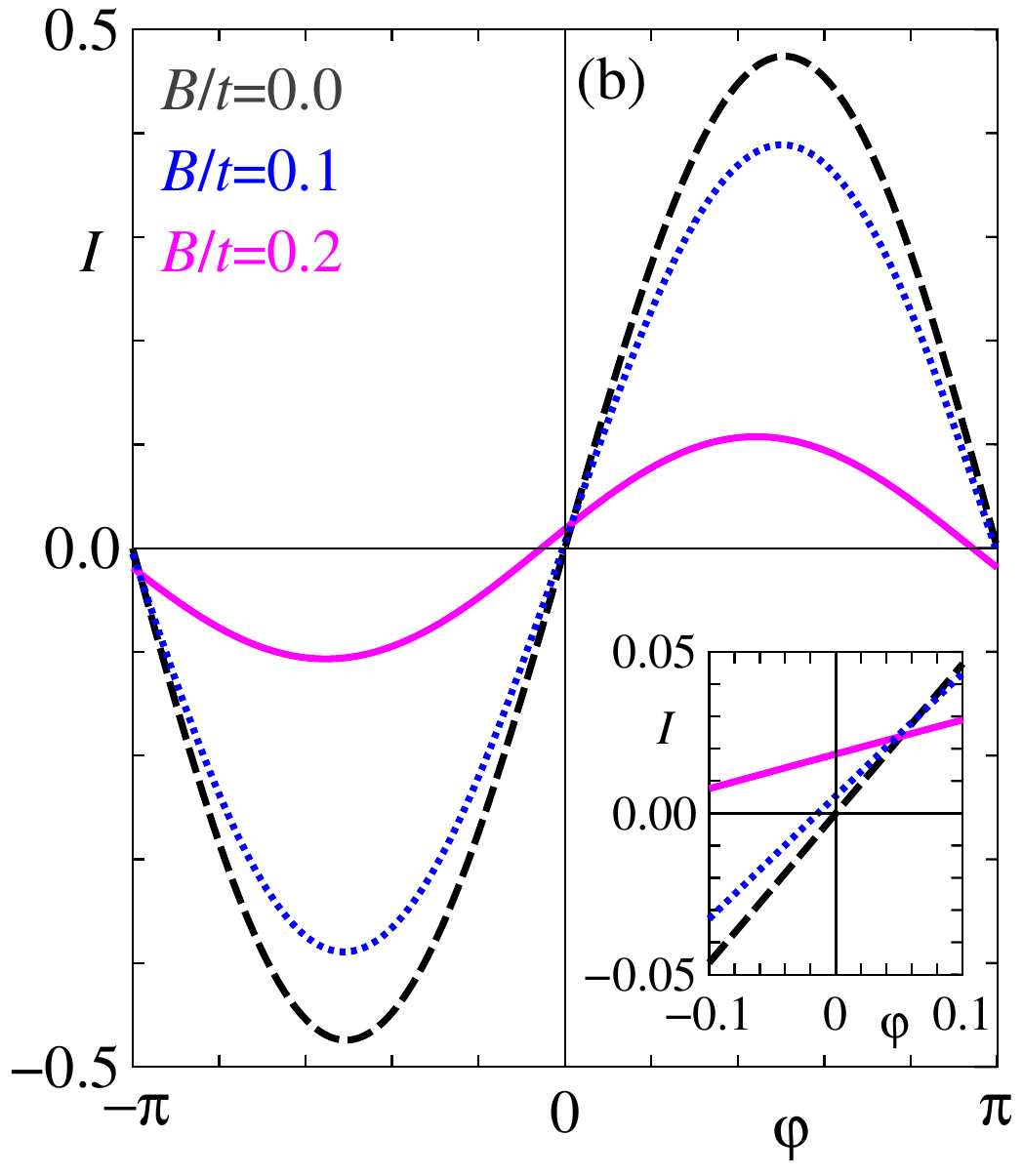}
\end{center}
\end{minipage}
\end{tabular}
\caption{
Numerical results of (a) $S_{z}^{\rm T}(\varphi)$
with the dashed line and solid line representing
$B/t = 0.0$ and $0.2$, respectively,
and (b) $I(\varphi)$ with the dashed line, dotted line, and solid line
representing $B/t = 0.0$, $0.1$, and $0.2$, respectively.
}
\end{figure}

The addition of the Zeeman field significantly enhances $S_{z}^{\rm T}$.
When $B/t = 0.1$ and $0.2$, $I(\varphi)$ is finite at $\varphi = 0$
and the point $\varphi_{0}$ at which $I(\varphi) = 0$
shifts to the negative direction.
This shows that the spin polarization generates the Josephson current
due to the inverse spin Hall effect.
Since the ground state energy is minimized when $I(\varphi) = 0$,
the system is most stable when the spin-induced Josephson current is
canceled by the contribution from the Josephson coupling.
In the case $B/t = 0.2$, $I$ is accurately expressed as
\begin{align}
    \label{eq:I_phi}
  I(\varphi) =  I_{1\rm s}\sin(\varphi-\varphi_{1\rm s})
                + \delta I_{2\rm s}\sin(2\varphi-\varphi_{2\rm s})
\end{align}
with $I_{1\rm s} \approx 0.1071$,
$\varphi_{1\rm s} \approx -0.1745$,
$\delta I_{2\rm s} \approx 0.0005711$,
and $\varphi_{2\rm s} \approx 0.4089$.
Note that $I(\varphi)$ is no longer antisymmetric with respect to $\varphi$.
This results in weak nonreciprocity between the critical currents
$I_{\rm c+}$ and $I_{\rm c-}$, which are defined as the maximum values of
$|I(\varphi)|$ in the regions $I > 0$ and $I < 0$, respectively.
The nonreciprocity is characterized by the diode efficiency defined as
\begin{align}
 \eta \equiv \frac{I_{\rm c+}-I_{\rm c-}}
                  {I_{\rm c+}+I_{\rm c-}} .
\end{align}
In the case $B/t = 0.2$, $I_{\rm c+} \approx 0.1075$ and
$I_{\rm c-} \approx 0.1067$ give $\eta \approx 0.003$,
which is small but finite.~\cite{comment2}
This indicates that the diode effect can appear in the tunneling regime,
although the regime is not ideal for observing it.
Equation~(\ref{eq:I_phi}) shows that $\eta$ is finite
only when $\delta I_{2\rm s} \neq 0$
and $\varphi_{2\rm s} \neq  2\varphi_{1\rm s}$ (${\rm mod}\, \pi$).
The asymmetry of $I(\varphi)$ due to the second harmonics
is essential for the appearance of the nonreciprocity,
supporting the scenario for the superconducting diode effect.~\cite{Tatara}

\begin{figure}[btp]
\vspace{0mm}
\begin{tabular}{cc}
\begin{minipage}{0.42\hsize}
\begin{center}
\includegraphics[height=4.5cm]{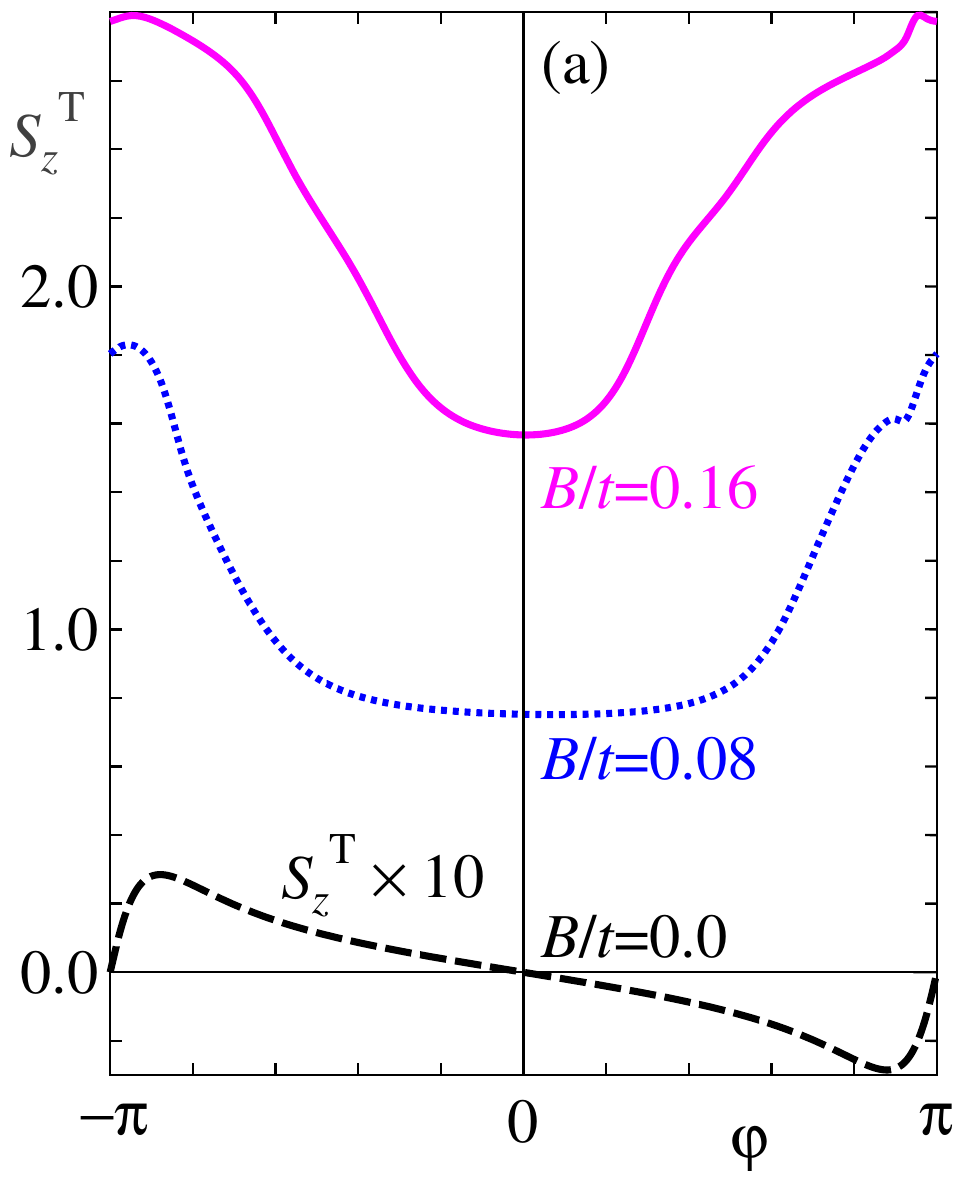}
\end{center}
\end{minipage}
\begin{minipage}{0.58\hsize}
\begin{center}
\hspace{0mm}
\includegraphics[height=4.5cm]{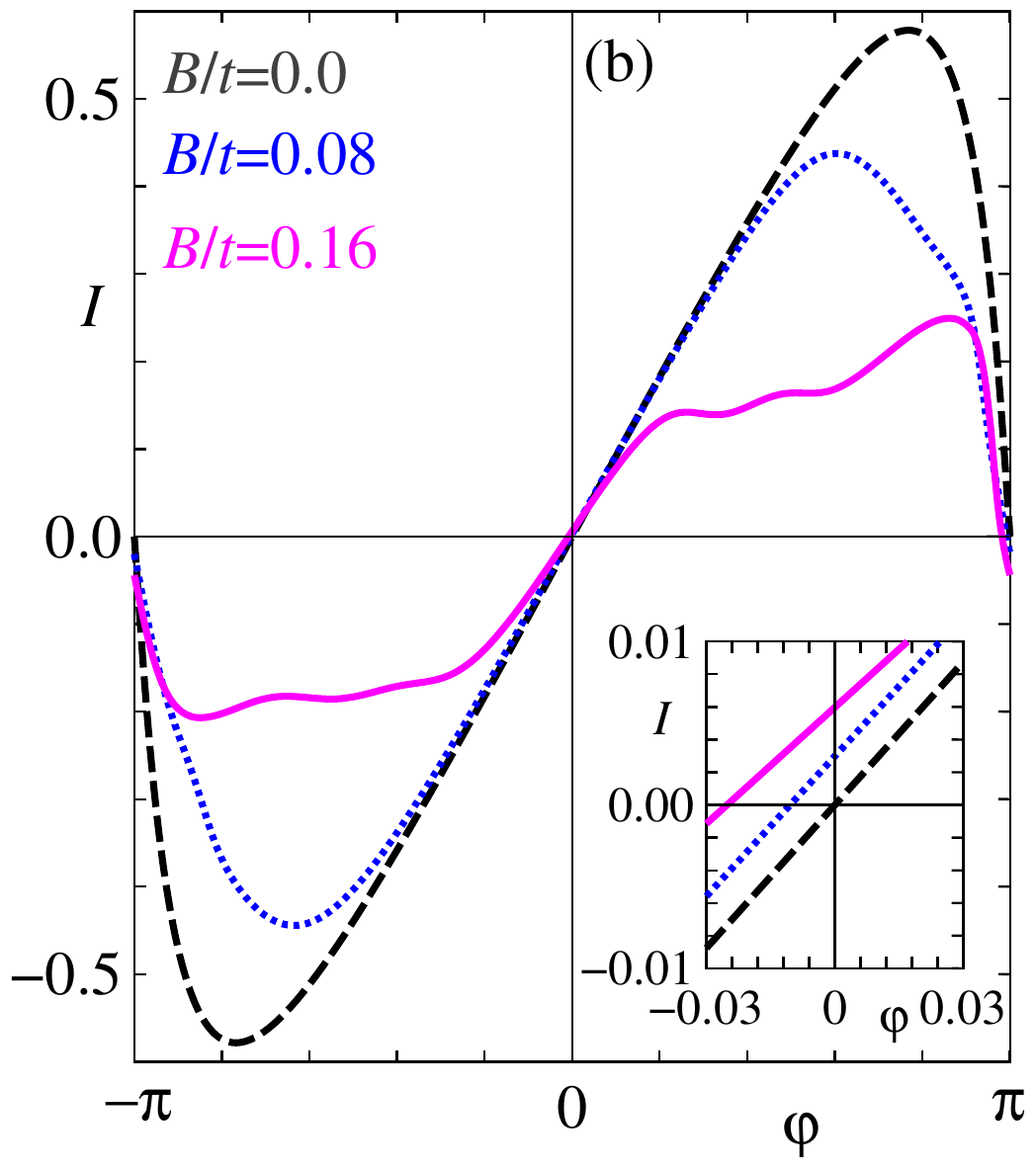}
\end{center}
\end{minipage}
\end{tabular}
\caption{
Numerical results of (a) $S_{z}^{\rm T}(\varphi)$ for
$B/t = 0.0$, $0.08$, and $0.16$ from bottom to top
(the dashed line at the bottom is displayed at 10 times its original scale)
and (b) $I(\varphi)$ with the dashed line, dotted line, and solid line
representing $B/t = 0.0$, $0.08$, and $0.16$, respectively.
}
\end{figure}
\begin{figure}[tpb]
\vspace{0mm}
\begin{tabular}{cc}
\begin{minipage}{0.5\hsize}
\begin{center}
\includegraphics[height=4.5cm]{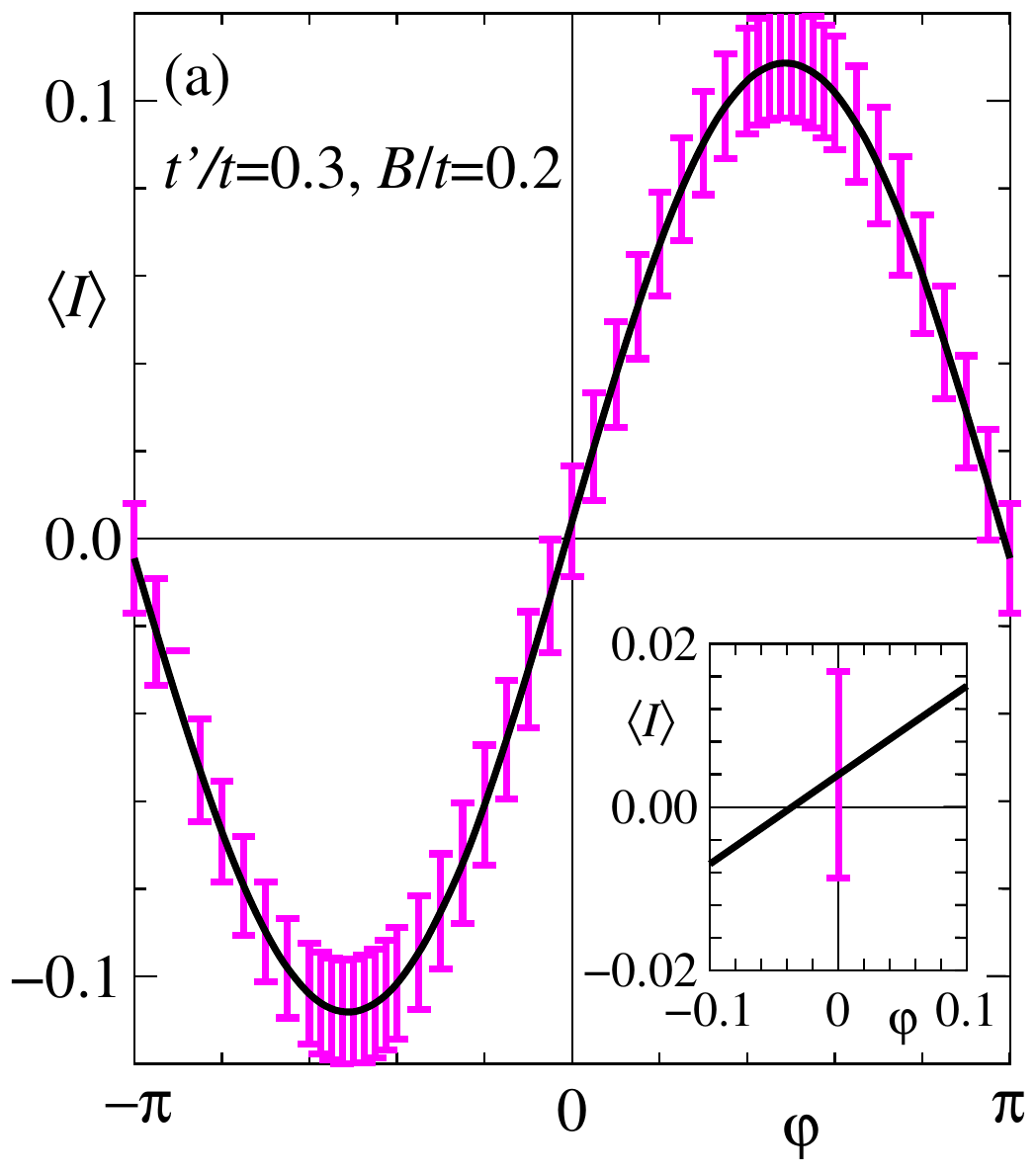}
\end{center}
\end{minipage}
\begin{minipage}{0.5\hsize}
\begin{center}
\hspace{0mm}
\includegraphics[height=4.5cm]{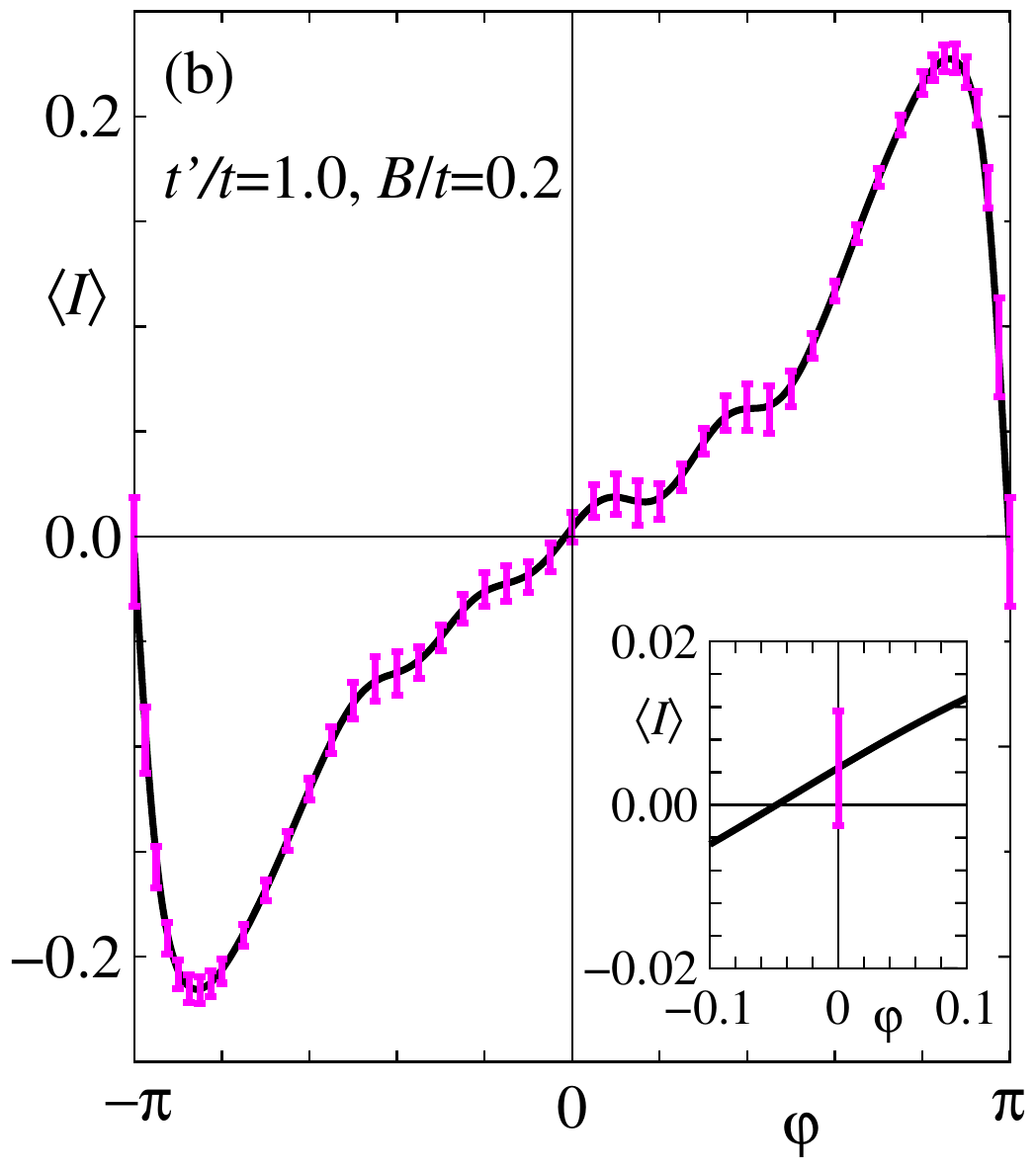}
\end{center}
\end{minipage}
\end{tabular}
\caption{
Numerical results of $\langle I(\varphi)\rangle$ with $B/t = 0.2$
in (a) the tunneling regime $t'/t = 0.3$
and (b) the metallic regime $t'/t = 1.0$.
}
\end{figure}
Let us next consider the metallic regime with $t' = t$
with a single impurity configuration.
Figures~3 shows (a) $S_{z}^{\rm T}(\varphi)$ and (b) $I(\varphi)$
in the cases $B/t = 0.0$, $0.08$, and $0.16$.
In the absence of the Zeeman field (i.e., $B = 0$),
we again observe $S_{z}^{\rm T}(\varphi) = -S_{z}^{\rm T}(-\varphi)$
in accordance with the antisymmetric feature of $I(\varphi)$,
showing the appearance of the spin Hall effect.
The presence of the Zeeman field significantly enhances $S_{z}^{\rm T}$
and causes $\varphi_{0}$ to deviate from  zero,
showing that the spin polarization generates the Josephson current
due to the inverse spin Hall effect.
In the case $B/t = 0.16$, $I_{\rm c+} \approx 0.249$ and
$I_{\rm c-} \approx 0.207$ give $\eta \approx 0.092$,
which is comparable to experimentally detectable values.
Apparent deviation of $I(\varphi)$ from a simple sinusoidal form
demonstrates that higher harmonics,
$\delta I_{n\rm s}\sin(n\varphi-\varphi_{n{\rm s}})$ with $n \ge 2$,
significantly contribute to $I(\varphi)$.
This indicates that the diode effect is pronounced
owing to an enhancement of the higher harmonics.

Figure~4 shows the ensemble average of the Josephson current,
$\langle I \rangle$, over 100 different impurity configurations
with $B/t = 0.2$ in the tunneling regime $t'/t=0.3$ and
the metallic regime $t'/t = 1.0$.
The error bar of each data point represents $\langle I \rangle \pm \Delta I$,
with $\Delta I$ denoting the standard deviation.
In the tunneling regime, $\langle I \rangle_{\rm c+} \approx 0.1086$ and
$\langle I \rangle_{\rm c-} \approx 0.1081$ give $\eta \approx 0.002$.
In the metallic regime, $\langle I \rangle_{\rm c+} \approx 0.228$ and
$\langle I \rangle_{\rm c-} \approx 0.216$ give $\eta \approx 0.027$.
These results show that the diode effect persists after the ensemble average
and is pronounced in the metallic regime more than in the tunneling regime.
Relatively large values of $\Delta I$ in the tunneling regime suggest
that simulations with large systems are needed to estimate $\eta$
quantitatively in this regime.

In this work, we demonstrated numerically that a magnetic field gradient can
induce a diode effect in a Josephson junction that accommodates
the spin Hall effect.
Our study corroborates a recent theoretical prediction in the tunneling regime,
and extends it to the metallic regime,
where the diode effect becomes quite substantial.
It is instructive to emphasize that typical diode effects
detected experimentally lie in the range of a few tens of percent.
Josephson junctions with strong spin-orbit coupled links, such as Nb/Pt/Nb,
could be a powerful platform to test the present prediction.

\begin{acknowledgment}

A.M. acknowledges support from the France 2030 government investment plan
managed by the French National Research Agency
under grant reference PEPR SPIN -- [SPINTHEORY] ANR-22-EXSP-0009
and [SUPERSPIN] ANR-24-EXSP-0012. 
G.T. thanks support by JSPS KAKENHI
[Grant-in-Aid for Scientific Research (B) 26K00658].

\end{acknowledgment}


\begin{thebibliography}{99}
\bibitem{comment1} Let us assume that the system contains impurities
that possess a rotationally invariant potential.
Under spatial inversion, the SOI due to such an impurity does not change
its local profile, while the spatial configuration
of impurities changes globally.
This means that the spatial inversion operation merely alters
the configuration of impurities.
\bibitem{comment2} Note that $\eta$, which varies from sample to sample,
fluctuates significantly from positive to negative in the tunneling regime.
In the metallic regime, $\eta$ rarely becomes negative.
\end{thebibliography}
\end{document}